\documentclass[a4paper,11pt]{article}
\usepackage{jheppub}
\usepackage[T1]{fontenc}
\usepackage[caption=false,font=normalsize,labelfont=sf,textfont=sf]{subfig}
\usepackage{tcolorbox}
\usepackage{tikz}
\usetikzlibrary{arrows}

\title{Thermodynamic properties of Schwarzschild black hole in non-commutative gauge theory of gravity}

\author[a]{Abdellah Touati}
\author[a,1]{and Slimane Zaim\note{Corresponding author.}}

\affiliation[a]{Department of Physics, Faculty of Matter Sciences, University of Batna-1, Batna 05000, Algeria}

\emailAdd{abdellah.touati@univ-batna.dz}
\emailAdd{zaim69slimane@yahoo.com}

\abstract{In this paper, we use the non-commutative (NC) gauge theory of gravity to investigate the thermodynamic properties of a deformed Schwarzschild Black Hole (SBH). Our results present a new scenario of black hole evaporation. As a first step, we describe the Arnowitt-Deser-Misner (ADM) mass, the Hawking temperature, and the entropy of NC SBH. The non-commutativity removes the divergent behavior of temperature, and the result shows a difference in the pole-equator temperature. These corrections also reveal a new fundamental length at the Planck scale order, $\Theta \sim 10^{-35}\,\mathrm{m}$. In the last stage of evaporation, the NC correction exposes a remnant entropy $S^{\Theta}_0$ of the NC SBH with a minimal mass $\hat{m}_0$, where the non-commutativity prevents the black hole from evaporating more than this minimal mass. Then, the description of the heat capacity and the Helmholtz free energy of the deformed black hole shows the effect of the NC gauge theory on the thermodynamic stability and the phase transitions. Finally, we investigate the influence of the black hole pressure on the stability and the phase transition of SBH in NC space-time. In the final stage of this scenario, the remnant black hole is thermodynamically stable. In this study, we find that the NC parameter plays a role similar to the thermodynamic variable. The results show a second-order phase transition of NC SBH.}

\keywords{non-commutative geometry, gauge gravity, Schwarzschild black hole, thermodynamic proprieties.}

\begin{document}
\maketitle
\flushbottom

\section{Introduction}

In 1915, Albert Einstein proposed one of the most successful theories of gravity. The success of this theory lies in predicting experimental results with high precision, such as the periastron advance of Mercury's orbit, the bending of light, gravitational red-shift, etc. More recent examples of this success also include the prediction of gravitational waves that was observed in 2016 by the LIGO experiment \cite{gw1}, and the recent observation of a black hole for the first time \cite{eh1} in 2019. However, this theory has problems with the description of gravity at the quantum scale, which led to the emergence of many theories with the aim of describing quantum gravity.

The first mechanism that unifies quantum mechanics and gravity was proposed by Hawking (1975) in the context of the semi-classical approach near the event horizon allowing a black hole to emit radiation and thus evaporate \cite{hawking1,hawking3}. Similar to thermal black body radiation, the black hole can emit a spectrum of particles, which phenomenon is known as Hawking radiation. The thermodynamic properties of black holes are part of black hole phenomenology, which itself is part of a larger research topic \cite{hawking3,harms,vaz,haranas2,hansen1,jawad1,chen1}. The understanding of black hole physics connects quantum processes and gravity, i.e., a theory of quantum gravity (QG). However, Hawking demonstrated that the thermal spectrum of the SBH has a temperature $T_H=\hbar\,c ^3/(8\pi G k_B M)$, which is clearly a divergent quantity when $M \rightarrow 0$. Thus the final stage of a black hole evaporation is not clear and its explanation requires a theory of QG. However, this problem led to the emergence of many theories that study the effect of some quantum gravity models on black hole thermodynamics, such as the effect of gravity’s rainbow \cite{lutfugr1,lutfugr2,lutfurgquin1}, and quantum deformation effect which predicts a minimal length \cite{lutfugup1,lutfueup2,lutfueup3,lutfueup4,lutfueup5,lutfuegup6}. Moreover, the study of the thermal radiation of black holes surrounded by quintessence matter has important influences on the black holes physics, which is considered as a candidate for dark energy, and in this context L{\"u}tf{\"u}o{\u{g}}lu et al. considered the SBH surrounded by quintessence in diverse formalism such as, generalized uncertainty principle, modified dispersion relations and in the context of rainbow gravity, to investigate its impact on the usual thermodynamic quantities \cite{lutfuqu1,lutfuqu2,lutfurgquin1}.

The non-commutativity theory has emerged in the same context as the above models and is motivated by string theory \cite{seiberg1}, provides a theoretical framework to eliminate the Hawking divergence in the classical case by quantifying space-time itself. In the past 20 years there have been a lot of interesting studies on NC thermodynamic proprieties of black holes \cite{piero1,lopez1,nozari1,nozari2,myung1,chai2,mukhe1,zaim1,linares,Hassanabadi1}. NC theory assumes a special canonical commutation relation similar to that in quantum mechanics, and it applies to coordinates of the space-time itself
\begin{equation}
[x^{\mu},x^{\nu}]=i\,\Theta^{\mu\nu}\,,
\end{equation}
where $\Theta^{\mu\nu}$ is an anti-symmetric real matrix for which we take only space-space non-commutativity, that is $\Theta_{0i}=0$, due to the known problem with unitarity \cite{unitarity1,unitarity2}. The ordinary product of any two arbitrary functions $f(x)$ and $g(x)$ defined over this space-time is then replaced by the star (Moyal) product ``$*$'', defined by
\begin{equation}\label{eq:1.2}
(f*g)(x)=f(x)\,\exp\left(\frac{i}{2}\,\Theta^{\mu\nu}\overleftarrow{\partial_{\mu}}\,\overrightarrow{\partial_{\nu}}\right)g(x)\,.
\end{equation}
Here we use the star product and the Seiberg-Witten (SW) maps \cite{seiberg1} in order to describe the gauge theory of gravity in NC space-time.

In this paper, we investigate the expression of the event horizon arising from the $\hat{g}_{00}$ component of the metric. The NC correction to the event horizon is obtained at the second order in $\Theta$ and depends on the observation angle $\theta$. We find that the singularity at the origin is shifted by the non-commutativity of space-time to a finite radius $r=2m$. Additionally, we obtain a correction in terms of the NC parameter up to the second order for each of the ADM mass, the Hawking temperature, and the black hole entropy. The obtained results indicate that non-commutativity removes the divergent behavior of the temperature and leads to a difference in the pole-equator temperature of the black hole, which gives us an estimation of $\Theta$ of order of the Planck scale. These findings give us a new scenario for the black hole evaporation.

We also examine the behavior of heat capacity and discuss the stability of the NC black hole. We take into account the Helmholtz free energy in NC space-time to investigate the phase transition. We then include the pressure due to NC quantum corrections and find a quasi-swallowtail structure appearing in the NC space-time. The non-commutativity shows the inflection point at the critical value of $\Theta$ which means a second-order phase transition in this geometry.

This paper is organized as follows. In section 2 we use the star product between tetrad fields and SW maps to describe NC corrections to the metric field. In section 3 we study the thermodynamic properties of the NC black hole where the corrections to the ADM mass, the Hawking temperature, and the entropy up to second-order in $\Theta$ are determined. The heat capacity in NC space-time is also obtained. The Helmholtz free energy is considered in the description of the black hole thermal proprieties. We then show and discuss the phase transition for SBH in the presence of pressure in NC space-time. Finally, in the last section, we present our remarks and conclusions.

\section{Non-commutative gauge theory of a Schwarzschild black hole}\label{Sec2}

The gauge theory of gravity is a theory of general relativity in the de-Sitter $\mathrm{SO}(4,1)$ group on a commutative 4-dimensional metric with spherical symmetry in Minkowski space-time, proposed by G. Zet et al in Refs. \cite{zet1,zet2}. The action is invariant under ordinary Lorentz transformations, and the gauge fields of gravity in the $\mathrm{SO}(4,1)$ group are denoted by $e^{a}_{\mu}$, with $a=0,1,2,3$, for the tetrad fields and $\omega^{ab}_{\mu}(x)=-\omega^{ba}_{\mu}(x)$, with $[ab]=[01],[02],[03],[12],[13],[23]$, for the spin connection. In this theory, if we preserve the NC space-time and base it partly on implementing symmetries on a flat NC space-time, we obtain a NC gauge theory of gravity in curved space-time, wherein the ordinary Lorentz transformations via the SW maps induce the NC canonical transformations of the deformed fields. The NC action stays invariant \cite{chai1,haranas1,haranas2,mebarki1}. In this case, the deformed gauge fields of gravity are denoted by $\hat{e}^{a}_{\mu}(x,\Theta)$ and $\hat{\omega}^{ab}_{\mu}(x,\Theta)$.

We are interested in the SBH. Let us consider the following static metric with a spherical symmetry
\begin{equation}\label{eqt2.6}
ds^{2}=-A^2(r)\,dt^2+\frac{1}{A^2(r)}\,dr^2+r^2\left(d\theta^2+\sin^2\theta d\phi^2\right),
\end{equation}
where $A(r)$ is an arbitrary function of $r$. In the tetrad formulation of General Relativity, the tetrad components are related to the metric by the following relation
\begin{equation}\label{eqt2.7}
g_{\mu\nu}=e^{a}_{\mu}\,e_{a\nu}\,.
\end{equation}
From the general case, we take a particular form of non-diagonal tetrad fields satisfying the relation \eqref{eqt2.7} as follows
\begin{equation}
e^{a}_{\mu}=\left[\begin{array}{cccc}
A(r) & 0 & 0 & 0 \\
0 & \frac{1}{A(r)}\sin\theta \cos\phi & r \cos\theta \cos\phi & -r \sin\theta \sin\phi \\
0 & \frac{1}{A(r)}\sin\theta \sin\phi & r \cos\theta \sin\phi & r \sin\theta \cos\phi \\
0 & \frac{1}{A(r)}\cos\theta & -r \sin\theta & 0
\end{array} \right].	\label{eqt2.8}
\end{equation}
The non-commutative correction to this metric was calculated in Ref. \cite{abdellah1} using the SW maps and the Moyal product. The deformed tetrad fields $\hat{e}^{a}_{\mu}(x,\Theta)$ can be described as a power-series in $\Theta$ up to second-order, following the same approach outlined in Ref. \cite{cham1}, as
\begin{equation}\label{eqt2.26}
\hat{e}^{a}_{\mu}(x,\Theta)=e^{a}_{\mu}(x)-i\,\Theta^{\nu\rho}\,e^{a}_{\mu\nu\rho}(x)+\Theta^{\nu\rho}\,\Theta^{\lambda\tau}\,e^{a}_{\mu\nu\rho\lambda\tau}(x)+\mathcal{O}(\Theta^{3})\,,
\end{equation}
where
\begin{align}\label{eqt2.28}
e^{a}_{\mu\nu\rho}=&\frac{1}{4}[\omega^{ac}_{\nu}\partial_{\rho}e^{d}_{\mu}+(\partial_{\rho}\omega^{ac}_{\mu}+R^{ac}_{\rho\mu})e^{d}_{\nu}]\eta_{cd},\\
e^{a}_{\mu\nu\rho\lambda\tau}=&\frac{1}{32}\left[2\{R_{\tau\nu},R_{\mu\rho}\}^{ab}e^{c}_{\lambda}-\omega^{ab}_{\lambda}(D_{\rho}R_{\tau\mu}^{cd}+\partial_{\rho}R_{\tau\mu}^{cd})e^{m}_{\nu}\eta_{dm}\right.\notag\\
&\left.-\{\omega_{\nu},(D_{\rho}R_{\tau\nu}+\partial_{\rho}R_{\tau\nu})\}^{ab}e^{c}_{\lambda}-\partial_{\tau}\{\omega_{\nu},(\partial_{\rho}\omega_{\mu}+R_{\rho\mu})\}^{ab}e^{c}_{\lambda}\right.\notag\\
&\left.-\omega^{ab}_{\lambda}\left(\omega^{cd}_{\nu}\partial_{\rho}e^{m}_{\mu}+\left(\partial_{\rho}\omega_{\mu}^{cd}+R_{\rho\mu}^{cd}\right)e^{m}_{\nu}\right)\eta_{dm}+2\partial_{\nu}\omega_{\lambda}^{ab}
\partial_{\rho}\partial_{\tau}e^{c}_{\lambda}\right.\notag\\ &\left.-2\partial_{\rho}\left(\partial_{\tau}\omega_{\mu}^{ab}+R_{\tau\mu}^{ab}\right)\partial_{\nu}e^{c}_{\lambda}-\{\omega_{\nu},(\partial_{\rho}\omega_{\lambda}+R_{\rho\lambda})\}^{ab}\partial_{\tau}e^{c}_{\mu}\right.\notag\\
&\left.-\left(\partial_{\tau}\omega_{\mu}+R_{\tau\mu}\right)\left(\omega^{cd}_{\nu}\partial_{\rho}e^{m}_{\lambda}+\left((\partial_{\rho}\omega_{\lambda}+R_{\rho\lambda})\right)e^{m}_{\nu}\right)\eta_{dm}\right]\eta_{cb}\,,
\end{align}
and
\begin{align}
\{\alpha,\beta\}^{ab}=\left(\alpha^{ac}\beta^{db}+\beta^{ac}\alpha^{db}\right)\eta_{cd},\qquad &[\alpha,\beta]^{ab}=\left(\alpha^{ac}\beta^{db}-\beta^{ac}\alpha^{db}\right)\eta_{cd},\\
D_{\mu}R_{\rho\sigma}^{ab}=\partial_{\mu}R^{ab}_{\rho\sigma}+&\left(\omega_{\mu}^{ac}R^{db}_{\rho\sigma}+\omega_{\mu}^{bc}R^{da}_{\rho\sigma}\right)\eta_{cd}.
\end{align}
The complex conjugate $\hat{e}^{a\dagger}_{\mu}(x,\Theta)$ of the deformed tetrad field is obtained from hermitian conjugate of the relation \eqref{eqt2.26}, that is
\begin{equation}\label{eqt2.32}
\hat{e}^{a\dagger}_{\mu}(x,\Theta)=e^{a}_{\mu}(x)+i\,\Theta^{\nu\rho}e^{a}_{\mu\nu\rho}(x)+\Theta^{\nu\rho}\Theta^{\lambda\tau}e^{a}_{\mu\nu\rho\lambda\tau}(x)+\mathcal{O}(\Theta^{3})\,,
\end{equation}
and the real deformed metric is given by \cite{chai1}
\begin{equation}\label{eqt2.33}
\hat{g}_{\mu\nu}(x,\Theta)=\frac{1}{2}\left[\hat{e}^{a}_{\mu}*\hat{e}^{b\dagger}_{\nu}+\hat{e}^{a}_{\nu}*\hat{e}^{b\dagger}_{\mu}\right]\eta_{ab}\,.
\end{equation}

In this study we take only one case $r-\phi$, which is given by the following matrix for $\Theta^{\mu\nu}$
\begin{equation}
\Theta^{\mu\nu}=\left(\begin{matrix}
0& 0 & 0 & 0 \\
0& 0 & 0 & \Theta \\
0& 0 & 0 & 0 \\
0& -\Theta & 0 & 0
\end{matrix}
\right), \qquad \mu,\nu=0,1,2,3\,,\label{eqt2.34}
\end{equation}
where $\Theta$ is a real positive constant.

The non-zero components of the NC tetrad fields $\hat{e}^{a}_{\mu}$ were calculated in Ref. \cite{abdellah1}. We obtain the deformed Schwarzschild components $\hat{g}_{\mu\nu}$ using the definition \eqref{eqt2.33}, with a correction up to second order in $\Theta$. For the Schwarzschild solution $A(r) =\left (1-2m/r\right)^{\frac{1}{2}}$ we find
{\small
\begin{align}
-\hat{g}_{00}&=\left(1-\frac{2 m}{r}\right)+\Theta^{2}\left\{\frac{m\left(88m^2+mr\left(-77+15\sqrt{1-\frac{2m}{r}}\right)-8r^2\left(-2+\sqrt{1-\frac{2m}{r}}\right)\right)}{16r^4(-2m+r)}\right\}\sin^2\theta\notag\\
&+\mathcal{O}(\Theta^4),\label{eqt2.44}\\
\hat{g}_{11}&=\left(1-\frac{2 m}{r}\right)^{-1}+\Theta^{2}\left\{\frac{m\left(12m^2+mr\left(-14+\sqrt{1-\frac{2m}{r}}\right)-r^2\left(5+\sqrt{1-\frac{2m}{r}}\right)\right)}{8r^2(-2m+r)^3}\right\}\sin^2\theta\notag\\
&+\mathcal{O}(\Theta^{4}),\label{eqt2.45}\\
\hat{g}_{12}&=-\Theta^{2}\left\{\frac{m \left(-4m^2+r^2\left(\sqrt{1-\frac{2 m}{r}}-7\right)+m r \left(16-17 \sqrt{1-\frac{2 m}{r}}\right)\right)}{32 r (2 m-r)^3}\right\}\sin(2\theta)+\mathcal{O}(\Theta^4),	\label{eqt2.46}\\
\hat{g}_{22}&=r^{2}+\Theta^2\left\{\frac{-8 m^3-6 m^2 r \sqrt{1-\frac{2 m}{r}}+50 m^2 r-6 r^3 \sqrt{1-\frac{2m}{r}}+23 m r^2 \sqrt{1-\frac{2 m}{r}}-43 m r^2}{32 r (r-2 m)^2}\right.\notag\\
&\left.+\frac{\cos(2\theta)\left(8m^3+6m^2r\left(\sqrt{1-\frac{2m}{r}}+5\right)+2r^3\left(5-3\sqrt{1-\frac{2m}{r}}\right)+mr^2\left(13\sqrt{1-\frac{2 m}{r}}-37\right)\right)}{32r(r-2m)^2}\right.\notag\\
&\left.+\frac{10r^3}{32r(r-2m)^2}\right\}+\mathcal{O}(\Theta^4),\label{eqt2.47}
\end{align}
\begin{align}
\hat{g}_{33}&=r^{2}\sin^2\theta+\Theta^{2}\left\{\frac{2m^2r\left(74-9
\sqrt{1-\frac{2m}{r}}\right)+4r^3\left(5-3\sqrt{1-\frac{2m}{r}}\right)+mr^2\left(47\sqrt{1-\frac{2m}{r}}-97\right)}{32r(r-2m)^2}\notag\right.\\
&\left.+\frac{m\cos(2\theta)\left(68 m^2+r^2\left(17-11\sqrt{1-\frac{2 m}{r}}\right)+2mr\left(9 \sqrt{1-\frac{2 m}{r}}-34\right)\right)-68 m^3}{32r(r-2m)^2}\right\}\sin^2\theta\notag\\
&+\mathcal{O}(\Theta^4).\label{eqt2.50'}
\end{align}}
Taking the limit $\Theta\rightarrow 0$ we recover the commutative metric \eqref{eqt2.6}, and in the equatorial plane $\theta=\pi/2$ the diagonal form of the metric is identical to that found in Ref. \cite{abdellah1}.

Then, the metric \eqref{eqt2.6} in NC space-time takes the following form
\begin{equation}\label{eq:2.17}
d\hat{s}^{2}=-\hat{g}_{00}dt^{2}+\hat{g}_{11}dr^{2}+2\hat{g}_{12}drd\theta+\hat{g}_{22}d\theta^{2}+\hat{g}_{33} d\phi^{2}\,.
\end{equation}
Note that the space-time metric \eqref{eqt2.6} is static and spherically symmetric, whereas the NC space-time metric \eqref{eq:2.17} is static but not spherically symmetric due to the presence of the cross product term $drd\theta$. This implies the existence of a coupling between $r$ and $\theta$ induced by non-commutativity in the plane of rotation that disappears when the non-commutativity goes to zero.

In our choice of non-commutative tensor $\Theta^{\mu\nu}$, the rotational invariance is broken but the translational invariance remains \cite{Sheikh-Jabbari:2008ybm}. The broken rotational invariance occurs in the $(r,\theta)$ plane in spherical coordinates. Thus the metric is non-diagonal in the presence of non-commutativity leading to only one non-zero off-diagonal term, $g_{12}$. We note that we have chosen non-commutativity between just two coordinates, but we could have equally chosen it to be between more than two at the cost of complicated calculations.

The NC space-time \eqref{eq:2.17} is independent of $t$ and $\varphi$, corresponding two killing vectors, the time translation and the azimuthal killing vectors. Then the NC space-time \eqref{eq:2.17} has two surfaces, a static limit surface and horizons. The static limit surface is one that can be found by solving $\hat{g}_{00}=0$, for which we find
\begin{equation}
r_{sls}^{\mathrm{NC}}=r_{h}\left[ 1+\left( \frac{\Theta }{r_{h}}\right) \left( \frac{4\sqrt{5}+1}{32\sqrt{5}}\right) \sin\theta +\left( \frac{\Theta }{r_{h}}\right) ^{2}\left( \frac{10+\sqrt{5}}{128}\right)\sin^2\theta\right],\label{eqt2.23}
\end{equation}
where $r_{h}=2m$ is the event horizon of the SBH in a commutative space-time. The space-time \eqref{eq:2.17} has a coordinate singularity at $\frac{1}{\hat{g}_{11}(r^{NC}_h)}=0$ corresponding to the event horizon
\begin{equation}
r_{h}^{NC}=r_{h}\left[ 1+\frac{3}{8}\left( \frac{\Theta }{r_{h}}\right) ^{2}\sin^2\theta \right].
\end{equation}

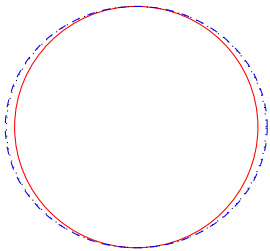
\begin{figure}[ht]
\centering
\begin{tikzpicture}[scale=0.8]
\draw [red] (0,0) circle (2cm);
\draw [ black, dotted] (0,0) ellipse (2.1355 cm and 2cm);
\draw [ blue, dashed] (0,0) ellipse (2.15266 cm and 2cm);
\end{tikzpicture}
\caption{\label{fig1}A schematic picture of the Schwarzschild black hole in the NC space-time. The red solid line represents the commutative Schwarzschild event horizon and the black dotted line represents the NC event horizon and the blue dashed line represents the static limit surface, with $\Theta=0.85$.}
\end{figure}

This solution has three parameters $r_{h}$, $\Theta$, and the observation angle $\theta$, while in the commutative space-time there is only one parameter $r_{h}$. It is a natural result of the deformation of the spherical symmetry of the black hole due to the rotation created by non-commutativity in coordinate space. Notice that if we take the limit $\Theta \rightarrow 0$ we exactly recover the commutative result. The radius of the event horizon increases with $\Theta $ as well with the angle $\theta $, and when $\theta $ takes the value $\pi/2$ we find the upper bound of radius of the event horizon. However when $\theta$ takes the values $0$ or $\pi $ then the NC correction term becomes absent and the radius of the event horizon becomes equal to the radius of Schwarzschild black hole (see Fig. \ref{fig1}). This confirms that the non-commutativity parameter plays the role of angular momentum induced by the rotation of the black hole due to the NC geometry of coordinates.

\section{Thermodynamic properties of the NC Schwarzschild black hole}

Let us now consider the ADM mass of the NC black hole, which can be obtained up to second order in the NC parameter as $\hat{M}=r_{h}^{\mathrm{NC}}/2$ \cite{Hassanabadi1,Hassanabadi2}, which can be obtained by solving the Eq. $(\hat{g}_{11}(r^{NC}_h))^{-1}=0$ for $m$. We can express this mass as a function of the event horizon $r_{h}$, $\Theta $ and the angle $\theta $ as follows
\begin{align}
\hat{M} &=\frac{r_{h}}{2}+\frac{3}{16r_{h}}\,\Theta^2\,\sin^2\theta\label{eqt3.10}\notag \\
&=m+M^{\mathrm{NC}}(\theta,\Theta)\,,
\end{align}
where $m=r_{h}/2$ is the ADM mass of the commutative SBH and $M^{\mathrm{NC}}(\theta,\Theta)$ is the NC correction term that depends on the angle $\theta$. Thus the mass of the NC black hole is
\begin{equation}\label{eq:4.1}
\hat{m}=\frac{1}{\int_{0}^{2\pi }\int_{0}^{\pi } \sqrt{\hat{g}_{22}\ast \hat{g}_{33}}\,d\theta\,d\varphi }\int_{0}^{2\pi }\int_{0}^{\pi } \hat{M}\,\sqrt{\hat{g}_{22}\ast \hat{g}_{33}}\,d\theta\, d\varphi\,.
\end{equation}
Then, using the deformed metric expressions \eqref{eqt2.47} and \eqref{eqt2.50'}, the mass of the NC SBH up to second order in $\Theta$ is given by
\begin{equation}\label{eq:4.2}
\hat{m}=m+\frac{\Theta ^{2}}{8\,r_{h}}\,,
\end{equation}
which reduces to the Schwarzschild black hole mass in the absence of non-commutative geometry ($\Theta =0$).
\begin{figure}[ht]
\centering
\includegraphics[width=0.5\textwidth]{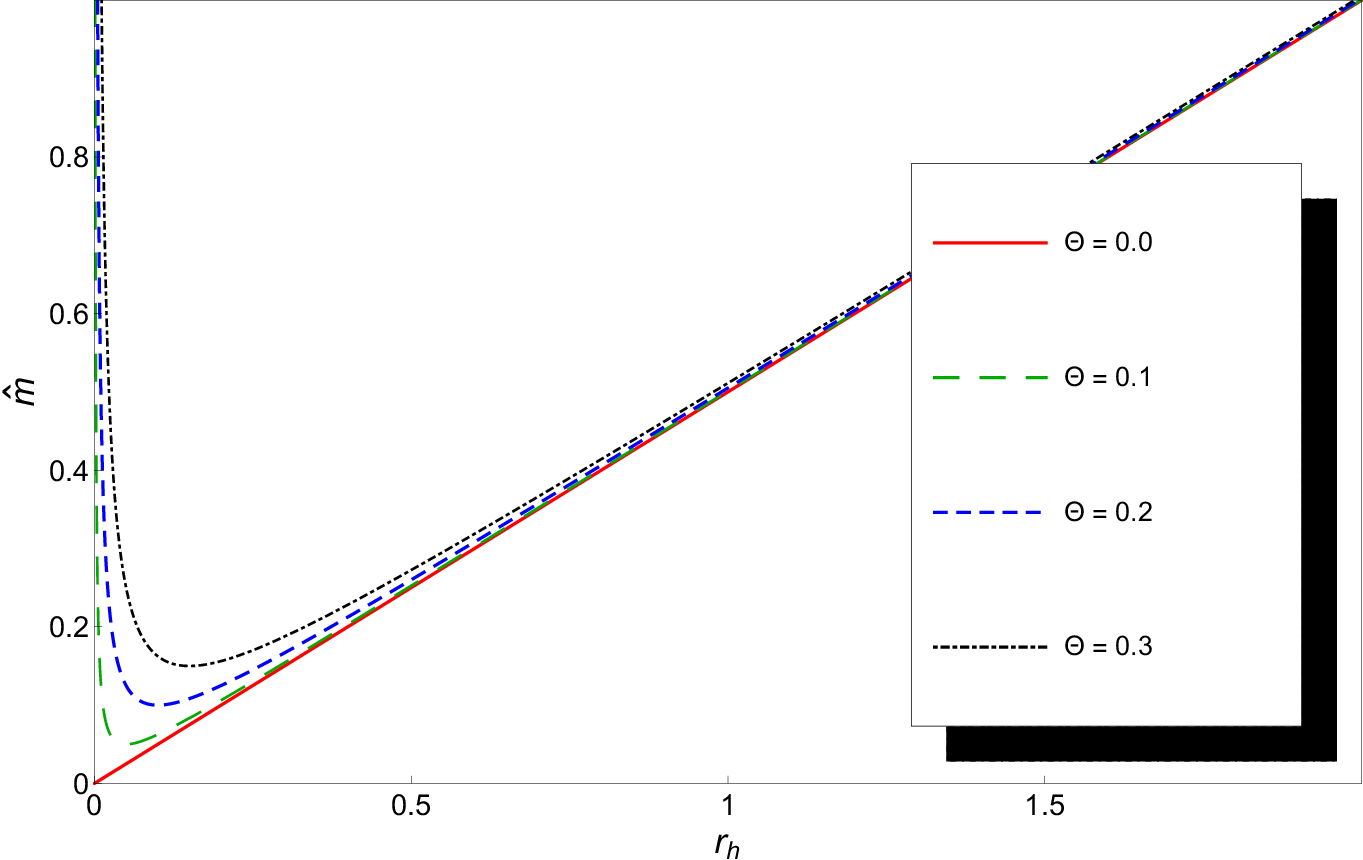}
\caption{Behavior of black hole mass as a function of $r_h$.}  \label{fig2}
\end{figure}

In Fig. \ref{fig2} we show the behavior of the black hole mass as a function of the horizon radius for different values of $\Theta$. In ordinary SBH, the mass has a linear behavior with the horizon radius $m=r_{h}/2$ and vanishes at the origin, while in NC space-time the mass of the NC SBH takes the minimal value $\hat{m}_{0}=0.5\,\Theta $ away from the origin at $r_{h}=r_{0}=0.5\,\Theta $ (in this case the NC parameter $\Theta $ plays the role of the mass, as was found in our previous work \cite{abdellah1}). This minimum mass increases with $\Theta $, meaning that non-commutative geometry creates a new minimum mass  preventing the mass from vanishing at the origin of the black hole. Thus if $r_h\rightarrow 0$ then $\hat{m}\rightarrow\infty$, and this corresponds to the singularity of the black hole at the origin. This is a clear confirmation of the general theory of relativity that links the geometry of space-time and mass (gravity) together.

\subsection{Hawking temperature}

The NC Hawking temperature can be obtained in the semi-classical framework using the NC surface gravity $\hat{\kappa}$ as
\begin{equation}
\hat{T}_H=\frac{\hat{\kappa}}{2\pi}=-\frac{1}{4\pi\sqrt{-\hat{g}_{00}(r,\Theta)*\hat{g}_{11}(r,\Theta)}}\left.\frac{\partial \hat{g}_{00}}{\partial r}\right|_{r=r_h^{\mathrm{NC}}},
\end{equation}
The NC temperature up to second order in the NC parameter $\Theta$, expanded up to $(1/r^4)$, can be written as follows
\begin{equation}\label{eq:4.4}
\hat{T}_H=\frac{1}{4\pi r_h}-\frac{3\Theta^2}{8 \pi r_h^3}\sin^2\theta\,,
\end{equation}

\begin{figure}[ht]
\centering
\includegraphics[width=0.55\textwidth]{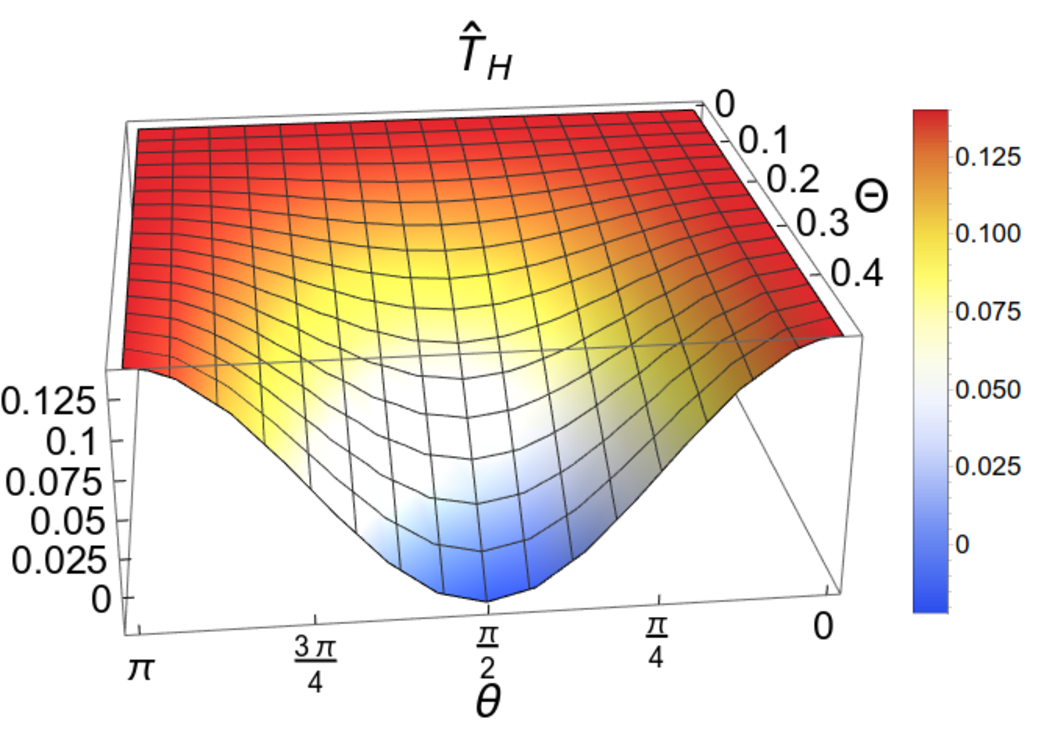}
\caption{Behavior of NC temperature distribution as a function of observation angle $\theta$ and NC parameter $\Theta$.}  \label{fig03}
\end{figure}

It is worth noting that the expression of the corrected temperature \eqref{eq:4.4} depends on the observation angle $\theta $, because the NC black hole is spherically asymmetric. We also note that the temperature of the NC black hole is the same as that in ordinary space-time when $\theta =0,\pi $, and this is actually a consequence of the choice of the NC parameter being perpendicular to the plane of rotation created by the non-commutativity of coordinates. The NC temperature depends on the observational angle $\theta $ and hence the radiation of the black hole is not the same in all directions of observation in NC space-time. The difference between poles-equator temperatures increases during the evaporation of the black hole. This result is similar to those obtained in some models and observations for the difference in temperature between the poles and equator of the Sun in the literature, e.g. \cite{poles1,poles2,poles3,poles4}. The difference in temperature between the poles and equator is obvious in the NC geometry, and it increases with $\Theta $ for small black holes as well (see Fig. \ref{fig03}).

In order to obtain an estimation of the NC  parameter $\Theta $, we use the temperature emitted from the surface horizon of the black hole, which can be calculated as follows
\begin{align}\label{eq:4.4'}
\hat{T}\left( r_{h}\right)  &=\frac{1}{\int_{0}^{2\pi }\int_{0}^{\pi } \sqrt{\hat{g}_{22}\ast \hat{g}_{33}}d\theta d\varphi }\int_{0}^{2\pi }\int_{0}^{\pi }  \hat{T}\left( r_{h},\theta\right)\sqrt{\hat{g}_{22}\ast \hat{g}_{33}}d\theta d\varphi \notag\\
&=\frac{1}{4\pi r_{h}}-\frac{\Theta^{2}}{4\pi r_{h}^{3}}\,.
\end{align}
Furthermore, we note that for $\Theta =0$ one recovers the Hawking temperature $T_H=\frac{1}{4\pi r_{h}}$ of the Schwarzschild black hole.

\begin{figure}[ht]
	\centering
	\includegraphics[width=0.5\textwidth]{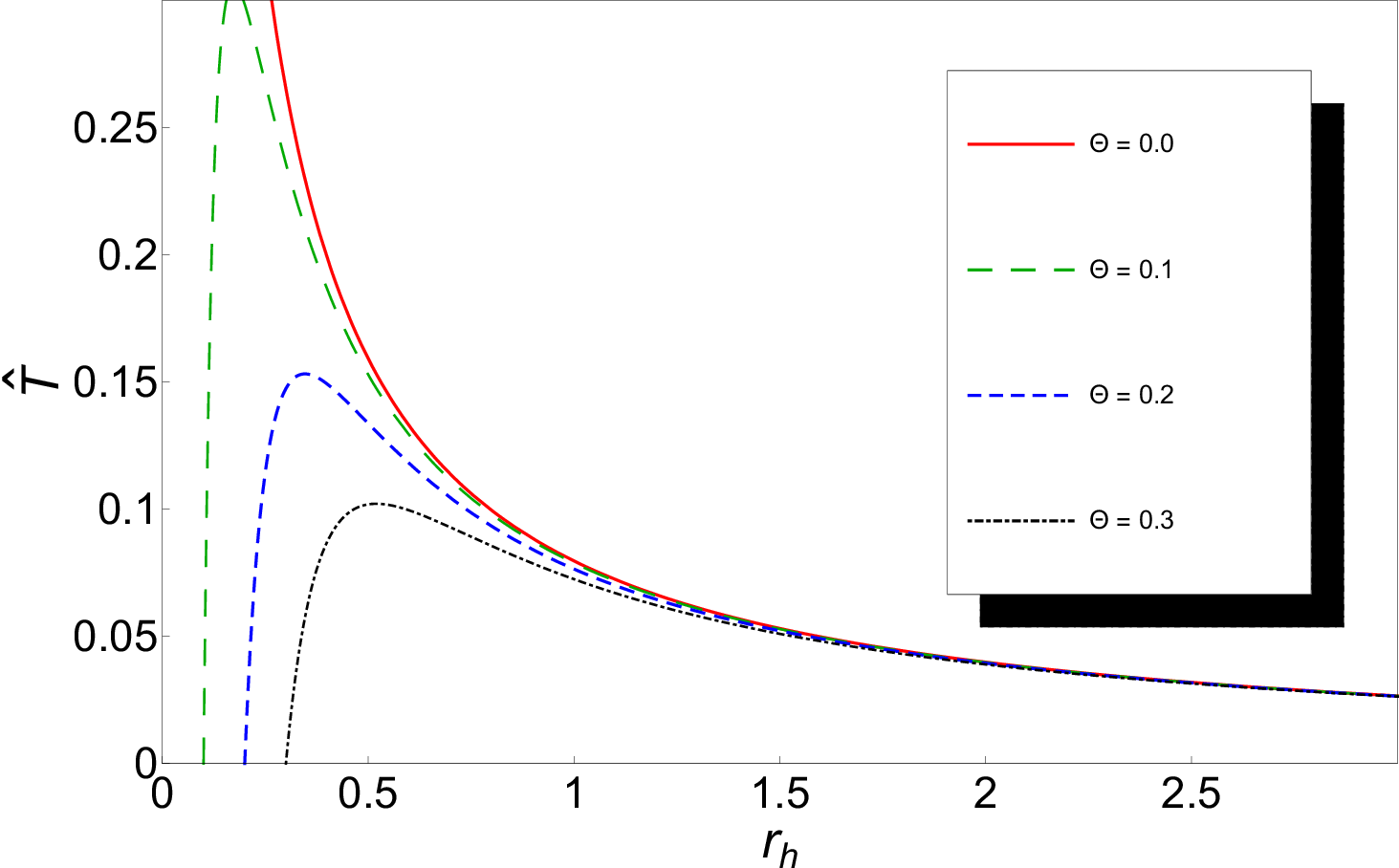}
	\caption{Behavior of NC temperature emitted from the surface horizon of the black hole as a function of $r_{h}$.}
	\label{fig3''}
\end{figure}

In Fig. \ref{fig3''} we show the behavior of Hawking temperature for the NC SBH as a function of the event horizon radius $r_{h}$ for various values of $\Theta $. The result shows that the non-commutativity removes the divergence of the Hawking temperature. The temperature of the NC SBH increases during its evaporation and reaches a maximum value $\hat{T}^{\max }=0.031/\Theta $ at the critical horizon radius $r_{h}^{c}=1.732\,\Theta $, then quickly falls to zero at the minimum horizon radius $r_{h}^{\min }=\Theta $. At this point, the black hole cannot radiate anymore, so there are no more processes of particle-antiparticle creation near the event horizon of the black hole with $r_{h}^{\min }$, which corresponds to the minimal mass $\hat{m}_0=r^{min}_h/2=0.5\,\Theta$.

The effect of non-commutativity is now clear. We observe that non-commutativity removes the divergence of the black hole radiation in its final stage of evaporation, similarly to the electric charge of Reissner-Nordrst\"{o}m black hole \cite{li}. In natural units system ($\hbar =k_{B}=c=1$), the thermal energy can be written as $E^{\mathrm{th}}=\hat{T}^{\max }$ and the mass of the black hole $\hat{m}=\frac{G}{2}r_{h}^c=0.866\,\Theta M_{\mathrm{Planck}}^{2}.$ The thermal energy and mass of the black hole should be of the same order of magnitude at the critical point $r_{h}^{C}$ for a significant result, and the NC parameter can be estimated as
\begin{equation}
\Theta\approx 1.523\times 10^{-35}m\sim l_{\text{Planck}}\,.
\end{equation}
This result is close to the one obtained in our previous work \cite{abdellah1,abdellah3}. Note also that our results obtained using the gauge theory do not exceed the Planck scale. However, there exist few papers that obtain a bound on the NC parameter $\Theta \sim10^{-1}l_{\text{Planck}}$ through the study of black-hole thermodynamics in NC spaces as in \cite{piero1,nicolini2,alavi,bound1}, and confirms that the NC property of space-time appears close to the Planck scale.

\subsection{Entropy}

We now turn to a second geometrical quantity, the NC SBH area, defined by
\begin{equation}\label{eq:4.5}
A_h^{\mathrm{NC}}=\int_{0}^{2\pi}\int_{0}^{\pi}\sqrt{\hat{g}_{22}*\hat{g}_{33}}\,d\theta\, d\phi\,.
\end{equation}
Using the distorted metric expressions \eqref{eqt2.47} and \eqref{eqt2.50'}, and taking the result up to second order only, we obtain the following result
\begin{align}\label{eq:4.6}
A_h^{\mathrm{NC}}=&2\pi\int_{0}^{\pi}\left[r^{\mathrm{NC}}_h+\frac{3r^{\mathrm{NC}}_h-m+\left(m+r^{\mathrm{NC}}_h\right)\cos(2\theta) }{16\,r^{\mathrm{NC}}_h}\right]\,\sin\theta \,d\theta,\notag\\
=&4\pi r_h^2+\frac{5\pi}{2}\Theta^2+\mathcal{O}(\Theta^4)\,.
\end{align}
The NC expression of entropy is related to the area of the NC black hole by the relation $\hat{S}=A_h^{NC}/4$. Thus we get
\begin{equation}\label{eq:4.7}
\hat{S}=\pi r_h^2+\frac{5\pi}{8}\Theta^2\,.
\end{equation}
When $\Theta=0$ the commutative entropy of the SBH is recovered $S=\pi r_h^2$.
\begin{figure}[ht]
\centering
\includegraphics[width=0.5\textwidth]{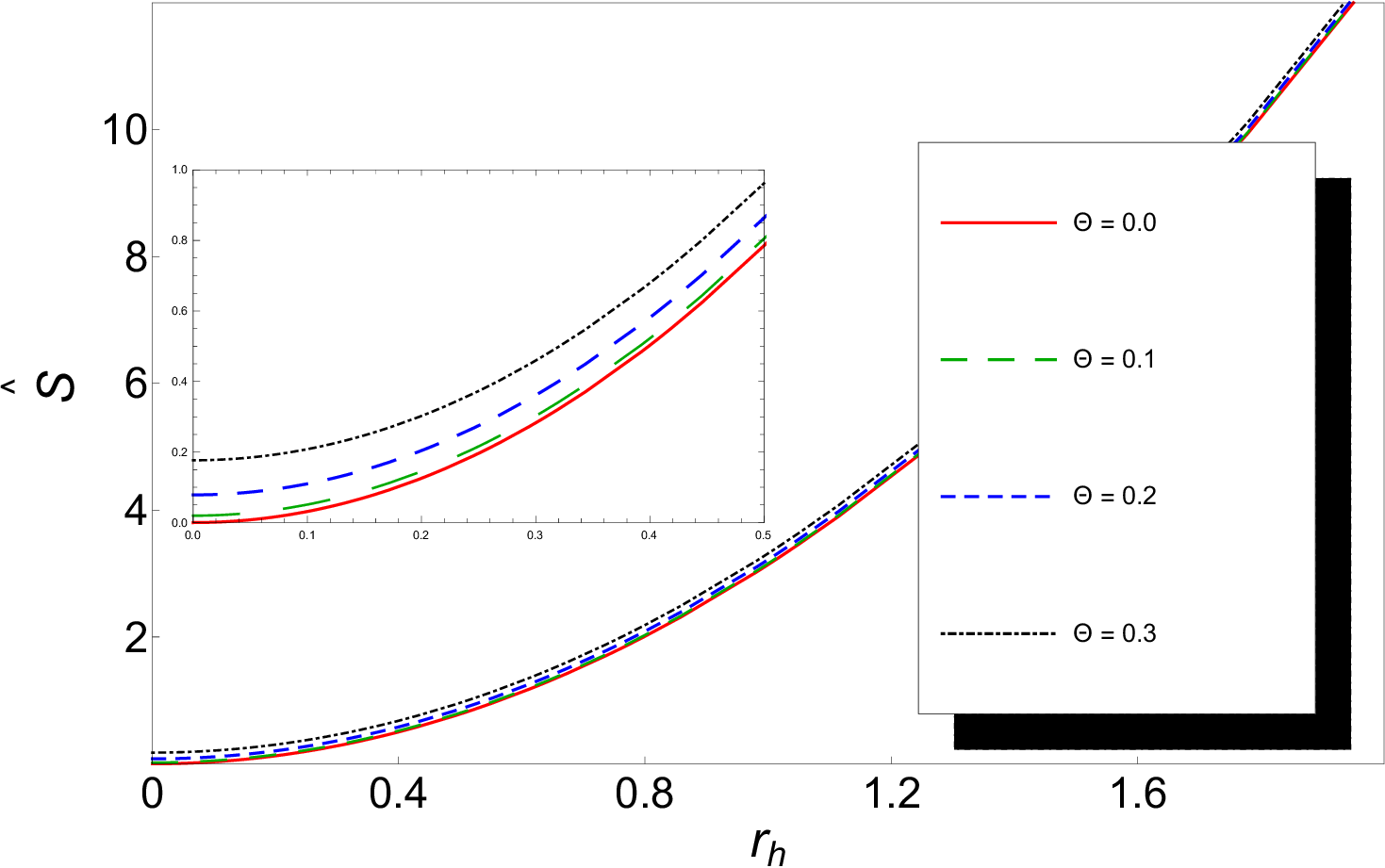}
\caption{Behavior of NC entropy as a function of $r_h$.}  \label{fig5}
\end{figure}

In Fig. \ref{fig5}, we observe a significant difference in the entropy between the commutative and NC space-time for smaller black holes, while the difference is insignificant for larger ones. At the final stage of evaporation and even after that we have a non-zero entropy \eqref{eq:4.8}, and this can be explained by the quantum structure of space-time
\begin{equation}\label{eq:4.8}
\lim_{r_h \rightarrow 0}\hat{S}=S_{0}^{\Theta }\approx \frac{5\pi}{8}\Theta^2.
\end{equation}
This expression emerges from pure quantum effects which originate from the NC propriety of space-time itself. The remnant entropy indicates a previously existing evaporated black hole, since the entropy is related to the area of the black hole via $S_0^{\Theta }=A_0^\Theta/4$. Thus, after evaporation the black hole still has an quantum area corresponding to quantum entropy. This strange quantum object has a quantum area and entropy $S_{0}^{\Theta }$ with a minimal mass $\hat{m}_0=0.5\,l_p$ and event horizon $r_{h }^{min}= l_{P}$, which can be seen as a microscopic remnant black hole.

The NC black hole is a thermodynamic system and hence obeys the first law of thermodynamics
\begin{equation}\label{eq:4.9}
d\hat{m}=\hat{T}d\hat{S}\,.
\end{equation}
The temperature $\hat{T}$ reads
\begin{align}\label{eq:4.9'}
\hat{T} &=\left(\frac{\partial \hat{m}}{\partial r_{h}}\right)\left(\frac{\partial \hat{S}}{\partial r_{h}} \right)^{-1} \notag \\
&=\frac{1}{4\pi r_{h}}-\frac{\Theta ^{2}}{16\pi r_{h}^{3}}\,.
\end{align}
We note that $\hat{T}$ in equation \eqref{eq:4.9'} calculated by the first law is not the same as $\hat{T}$ in equation \eqref{eq:4.4'} calculated by surface gravity. In the NC space-time, the mass of a black hole contains a term associated with a non-commutativity parameter. The classical form of the first law is modified by an extra factor
\begin{equation}\label{eq:4.9''}
c\left( r_{h},\Theta \right) d\hat{m}=\hat{T}d\hat{S}\,,
\end{equation}
where $c\left( r_{h},\Theta \right) $ is
\begin{equation}
c\left( r_{h},\Theta \right) =1-\frac{3\Theta ^{2}}{4r_{h}^{2}}\,.
\end{equation}
Substituting these values in Eq. \eqref{eq:4.9''} for $c\left( r_{h},\Theta \right) $, the area law is restored.

\subsection{Heat capacity}

We now check the thermodynamic stability of the black hole by the signature of heat capacity in NC space-time
\begin{align}\label{eq:4.10}
\hat{C}&=\hat{T}\left(\frac{\partial\hat{S}}{\partial \hat{T}}\right)=\hat{T}\left(\frac{\partial\hat{S}}{\partial r_h}\right)\left(\frac{\partial\hat{T}}{\partial r_h}\right)^{-1}\notag\\
&=-2\pi r_h^2\,\frac{(r_h^2 - \Theta^2)}{(r_h^2 - 3 \Theta^2)}\,,
\end{align}
using the Hawking temperature (Eq. \eqref{eq:4.4}) and the entropy (Eq. \eqref{eq:4.7}).

\begin{figure}[ht]
\centering
\includegraphics[width=0.5\textwidth]{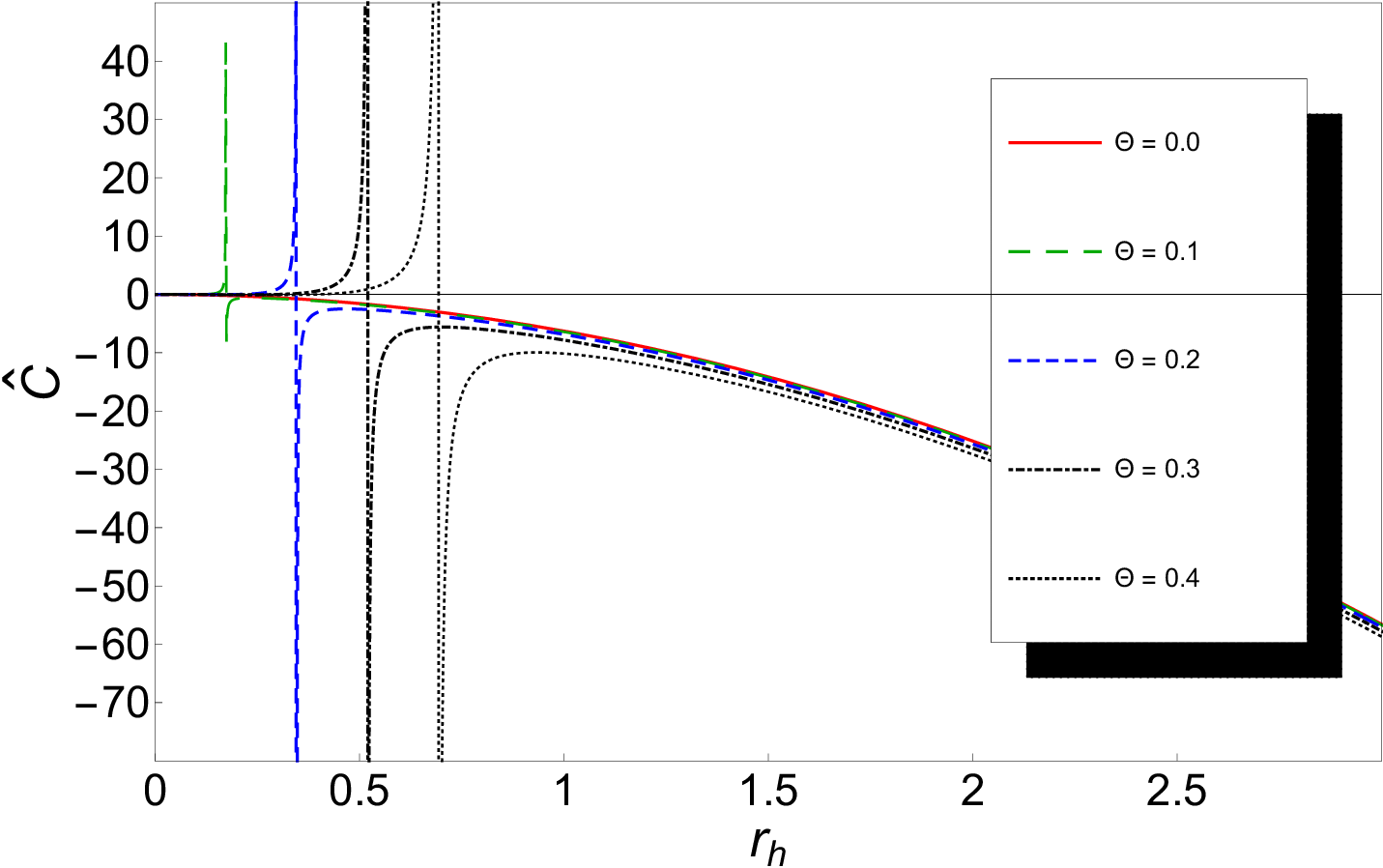}
\caption{Behavior of black hole heat capacity as a function of $r_h$.}  \label{fig7}
\end{figure}

We plot in Fig. \ref{fig7} the behavior of the heat capacity of the SBH in NC space-time. We observe that for $\Theta=0$ we recover the standard heat capacity, which is always negative, hence a phase transition in the commutative SBH does not exist. From this figure we observe that the heat capacity is equal to zero at $r_h=r_h^{\min}$, which means that this black hole has one physical limitation point \cite{saheb1}, and has a divergence at $r_h=r^{\mathrm{crit}}_h$ (at this point the black hole reaches the maximum temperature). This implies one phase transition point.

In NC geometry, a larger (massive) black hole has a negative heat capacity for $r_h>r^{\mathrm{crit}}_h$, which corresponds to a non-equilibrium (unstable) system, while for a smaller one it has a positive heat capacity for $r_h^{\min}<r_h<r^{\mathrm{crit}}_h$, hence the system is in (stable) equilibrium. The divergence of heat capacity describes a phase transition of SBH in the NC space-time at a critical point $r_h=r^{\mathrm{crit}}_h$, where this point grows with $\Theta$, and the stable stage $r_h^{\min}<r_h<r^{\mathrm{crit}}_h$ with positive heat capacity increases with $\Theta$, which means that the SBH takes longer to stop radiating and evaporating.

\subsection{Helmholtz free energy}

For more details on the analysis of the phase transition we need to investigate the Helmholtz free energy, which is defined as the Legendre transform of the ADM mass
\begin{align}\label{eqt3.14}
\hat{F}&=\hat{m}-\hat{T} \hat{S},\notag\\
&=\frac{r_h}{4}+\frac{7\Theta ^{2}}{32r_{h}}+\mathcal{O}(\Theta^4).
\end{align}

\begin{figure}[ht]
	\centering
	\includegraphics[width=0.5\textwidth]{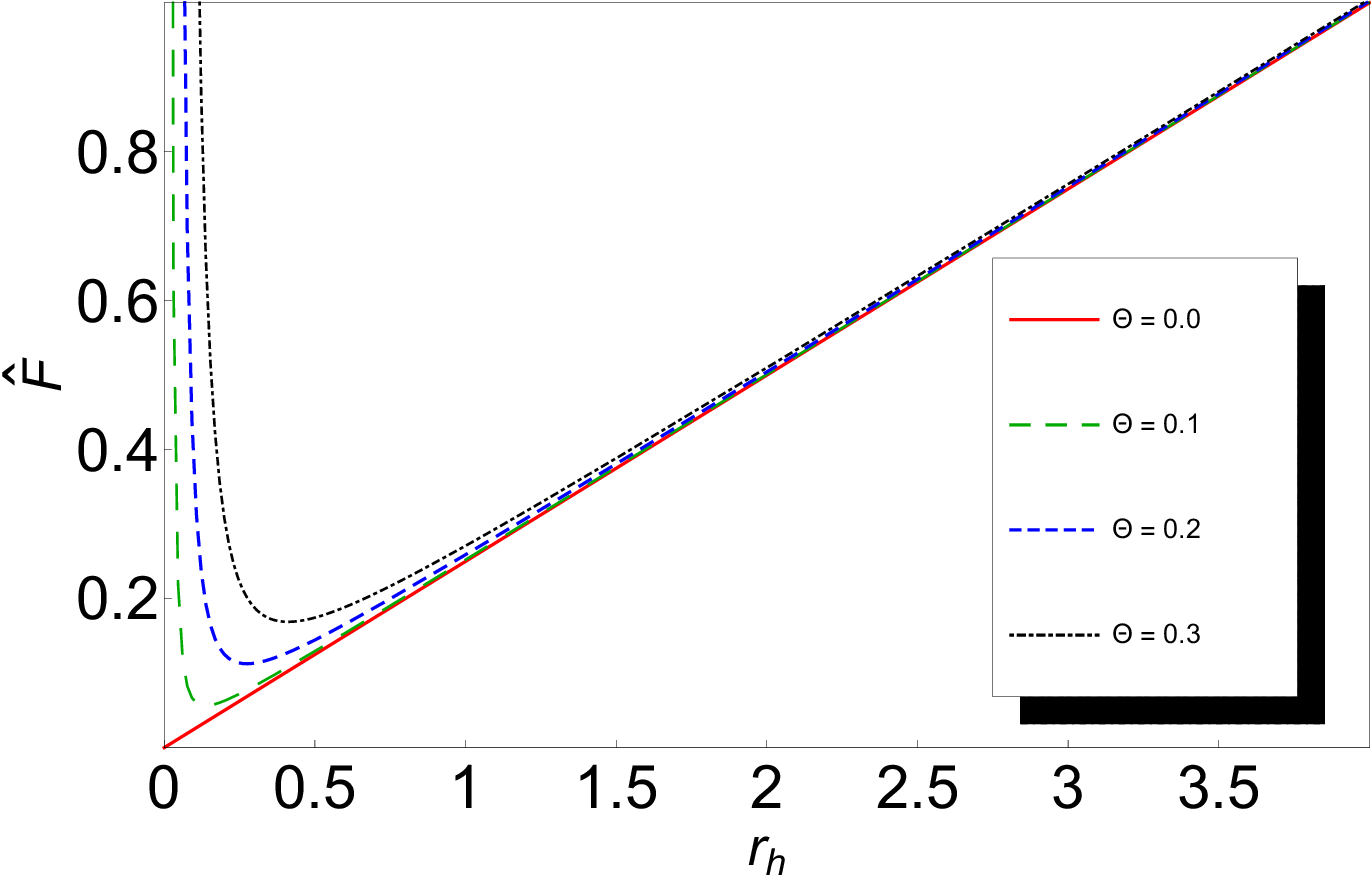}
	\caption{Behavior of Helmholtz free energy $\hat{F}$ as a function of $r_h$ in NC space-time.}  \label{fig8}
\end{figure}

In Fig. \ref{fig8}, we notice that the behavior of the Helmholtz free energy has a minimum in NC space-time given by
\begin{equation}\label{eqt3.15}
\frac{\partial \hat{F}}{\partial r_h}=0\,,
\end{equation}
where the location of this minimum increases with $\Theta$, which represents a stable region. This result confirms the one obtained in Fig. \ref{fig7}, which indicates that a smaller black hole is stable compared to a larger one.

\begin{figure}[ht]
	\centering
	\includegraphics[width=0.5\textwidth]{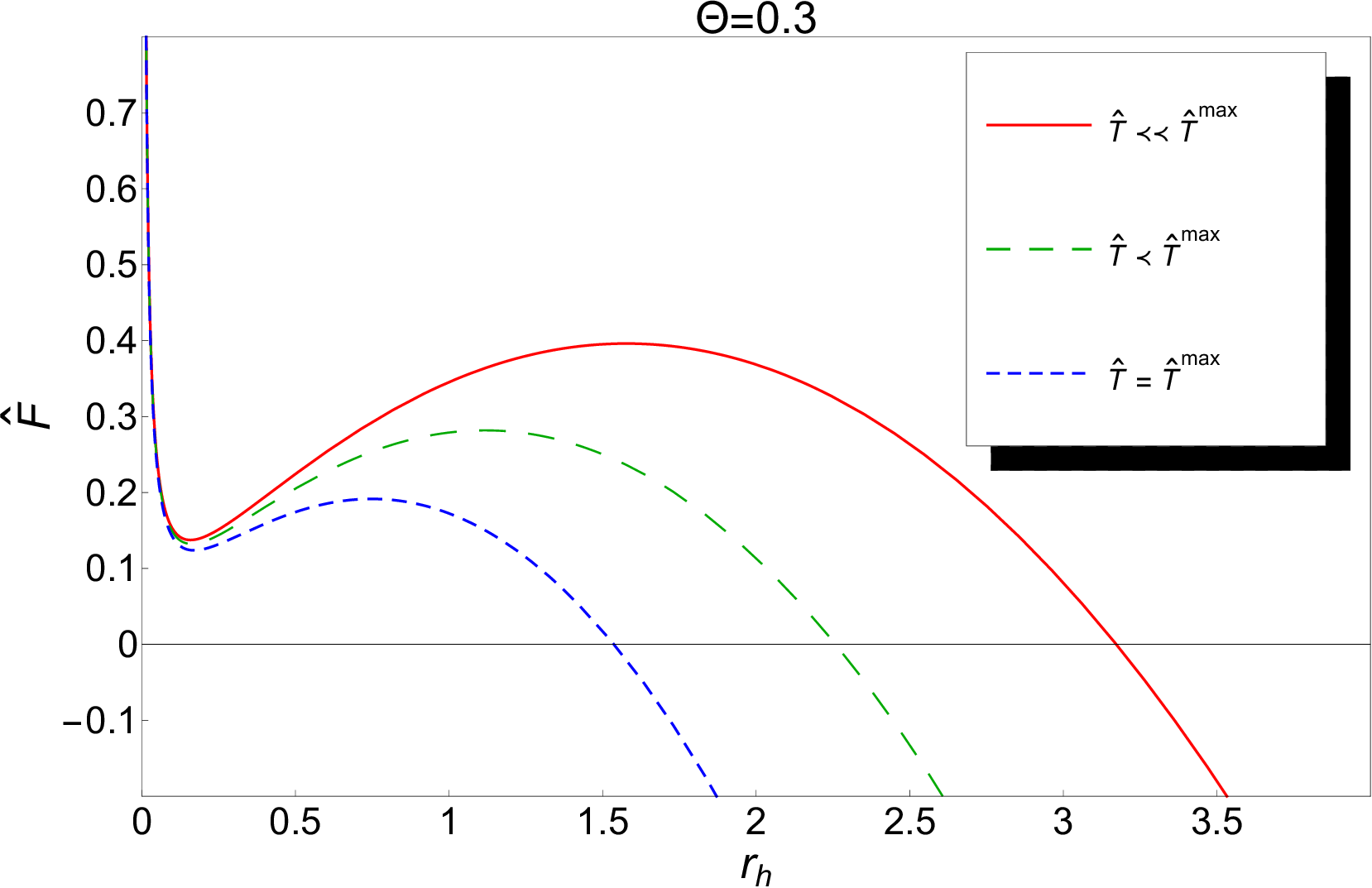}
	\caption{Behavior of Helmholtz free energy $\hat{F}$ as a function of $r_h$ in NC space-time for different Hawking temperatures $\hat{T}$.}  \label{fig9}
\end{figure}

The effect of Hawking temperature on the Helmholtz free energy is illustrated in Fig. \ref{fig9}, where we plot the following expression for different Hawking temperatures $\hat{T}$
\begin{equation}
\hat{F}=\frac{r_h}{2}-\pi r_h^2 \hat{T} + \Theta^2\left(\frac{1}{8r_h}-\frac{5\pi}{8}\hat{T}\right).
\end{equation}

It can be seen from Fig. \ref{fig9} that when the Hawking temperature of the black hole is taken into account, the Helmholtz free energy has two extremum (stable and unstable), which can be obtained by solving Eq. \eqref{eqt3.15}. From these behaviors, we conclude that a large black hole, corresponding to a maximum of the Helmholtz free energy, is thermodynamically unstable, while the stable one has a minimum Helmholtz free energy, contrary to the commutative SBH which is always unstable.
\begin{figure}[ht]
\centering
\includegraphics[width=0.5\textwidth]{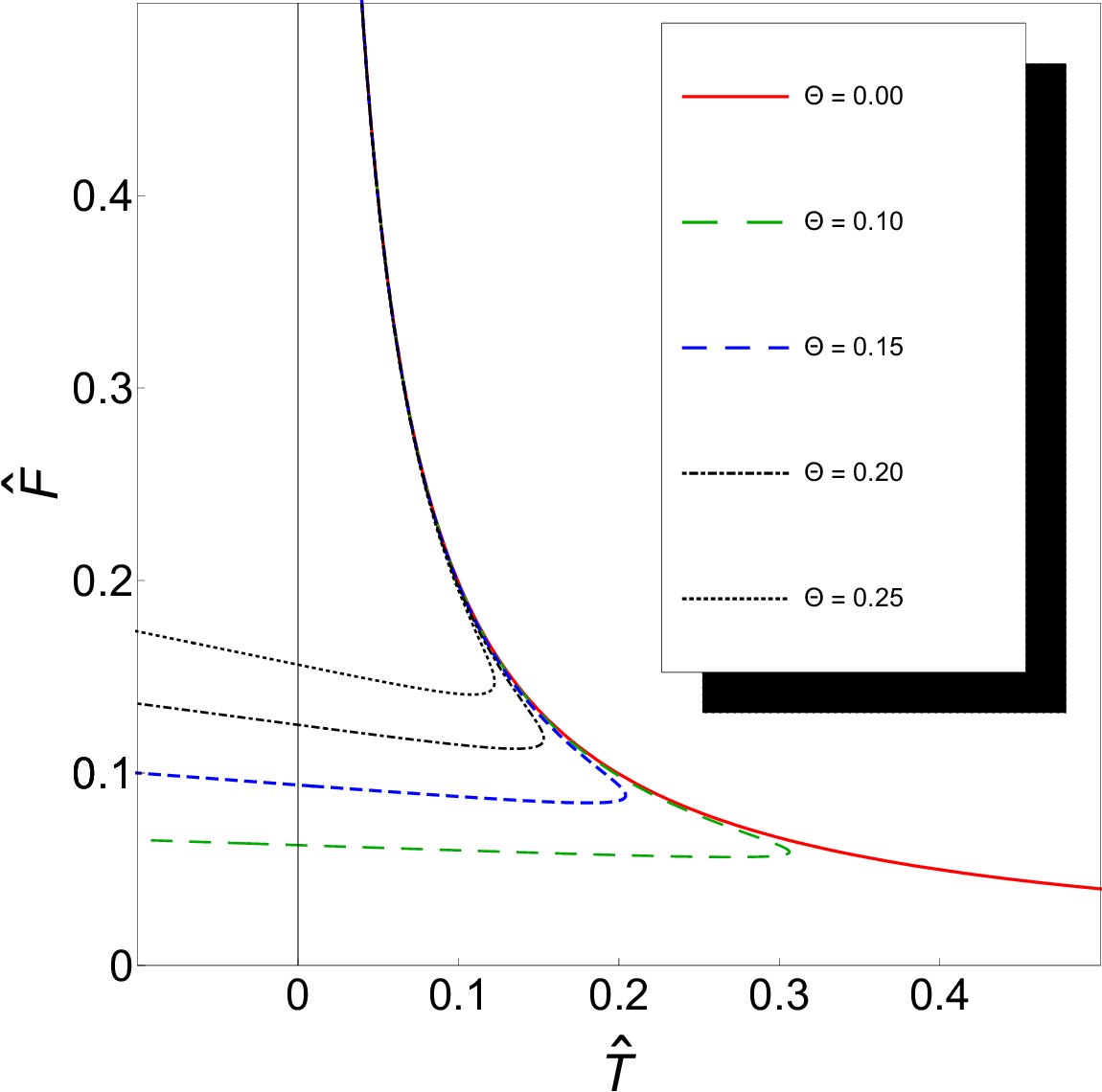}
\caption{Behavior of the Helmholtz free energy $\hat{F}$ as a function of NC Hawking temperature $\hat{T}$.}  \label{fig10}
\end{figure}

In Fig. \ref{fig10} we represent the behavior of the Helmholtz free energy as a function of the NC Hawking temperature with different values of the NC parameter $\Theta$. We observe that in the NC space-time the Helmholtz free energy decreases rapidly with increasing temperatures for small $\Theta$ (from infinity at $\hat{T} \sim 0$). Then it starts to increase slowly when the temperature bounces back from a maximum $\hat{T}^{\max}\approx 0.031/\Theta$, and becomes more rapidly increasing with $\Theta$. In the commutative case $\Theta=0$, the Helmholtz free energy decreases with Hawking temperature and does not effect its behavior. One can see that this behavior is similar to that obtained with the modified thermodynamic first law when the surface tension is considered as in Refs. \cite{chen1,jawad1,hansen1}. However, in our work, this behavior emerges from the quantum structure of space-time in the modified SBH by non-commutativity. Using these works as analogies, we can see non-commutativity as a space-time tension on the quantum scale.

\subsection{The black hole pressure}

Our next step is motivated by the recent papers \cite{calmet1,calmet2,delgado}. The authors of these papers suggest that quantum gravitational corrections to the thermodynamic proprieties of SBH lead to pressure in this black hole. Here we consider a quantum correction for SBH by the NC gauge theory. For that, we add the pressure of the black hole by ad-hoc $\hat{P}d\hat{V}$ term to the first law of thermodynamics \eqref{eq:4.9}, in order to analyze SBH phase transition in the NC space-time
\begin{equation}\label{eqt3.18}
d\hat{m}=\hat{T}d\hat{S}-\hat{P}d\hat{V}\,,
\end{equation}
where $\hat{V}$ denotes the total volume of SBH in NC space-time, which can be computed using the same step in \eqref{eq:4.1} (we stop at the second order in $\Theta$)
\begin{align}\label{eqt3.19}
\hat{V}&=\frac{4\pi}{3}\frac{1}{\int_{0}^{2\pi}\int_{0}^{\pi}\sqrt{\hat{g}_{22}*\hat{g}_{33}}\,d\theta\, d\phi}\int_{0}^{2\pi}\int_{0}^{\pi}\left(r^{\mathrm{NC}}_h\right)^3\sqrt{\hat{g}_{22}*\hat{g}_{33}}\,d\theta\, d\phi\notag\\
&=\frac{4\pi}{3}r_h^3 +\pi r_h \Theta^2+\mathcal{O}(\Theta^4)\,,
\end{align}
and its conjugate, the pressure $\hat{P}$
\begin{align}
\hat{P}&=-\left(\frac{\partial \hat{m}}{\partial \hat{V}}\right)=-\left(\frac{\partial \hat{m}}{\partial r_h}\right)\left(\frac{\partial \hat{V}}{\partial r_h}\right)^{-1}\notag\\
&=-\frac{1}{8\pi r_h^2}+\frac{\Theta^2}{16\pi r_h^4}+\mathcal{O}(\Theta^4)\,,
\end{align}
where $P=-1/(8\pi r_h^2)$ is the commutative term of the pressure.
\begin{figure}[ht]
\centering
\includegraphics[width=0.5\textwidth]{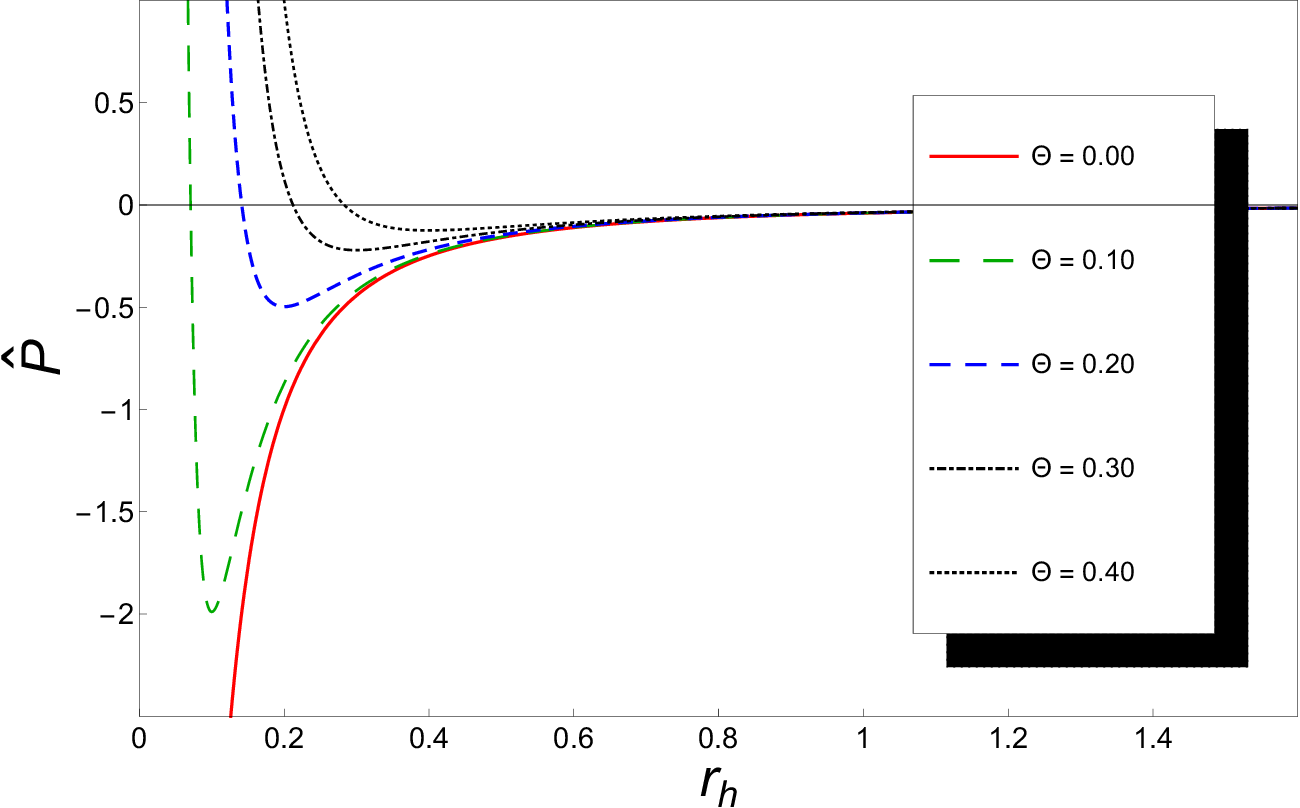}
\caption{Behavior of pressure $\hat{P}$ as a function of $r_h$.} \label{fig11}
\end{figure}

In Fig. \ref{fig11}, we show the behavior of the NC SBH pressure as a function of the event horizon radius, for different values of $\Theta $. The results indicate that the non-commutativity removes the divergence of the NC SBH pressure, which decreases rapidly until it reaches its minimum value at the point $\left( -0.02/\Theta ^{2},\Theta \right) $, and then increases to take a value of zero at  $r_{h}=0.71\Theta $. This means that the NC coordinates create a new minimum pressure, which is a transition point associated with the interaction of the black hole with NC space, where the effect of the pressure on the black hole induced by the non-commutative space-time increases with decreasing $r_{h}$ until the equilibrium point at $r_{h}=0.71\Theta$. Note that when $r_h$ is small, the pressure is negative and can be interpreted as the pressure exerted by the black hole on space-time. This pressure is induced by the radiation of the black hole.

\subsubsection{Gibbs free energy}

\begin{figure}[ht]
\centering
\includegraphics[width=0.5\textwidth]{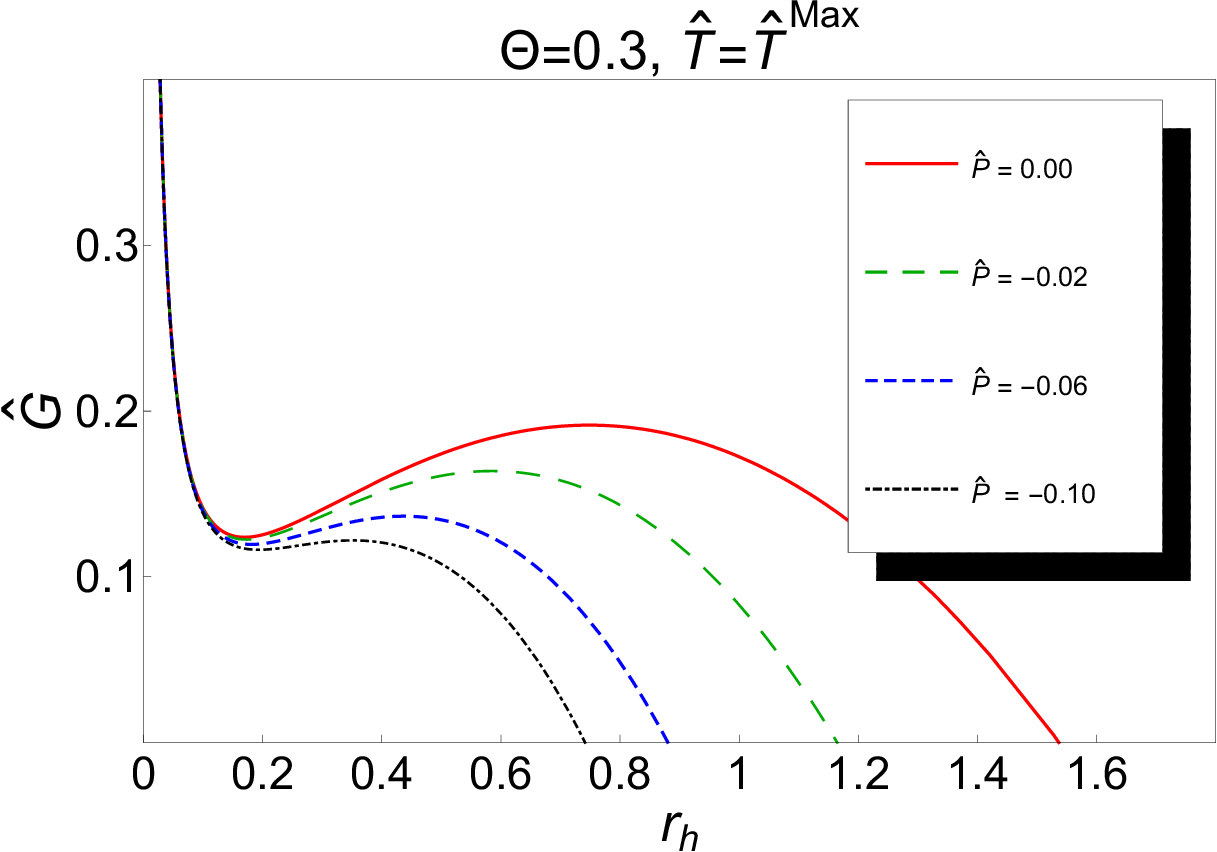}\hfill
\includegraphics[width=0.5\textwidth]{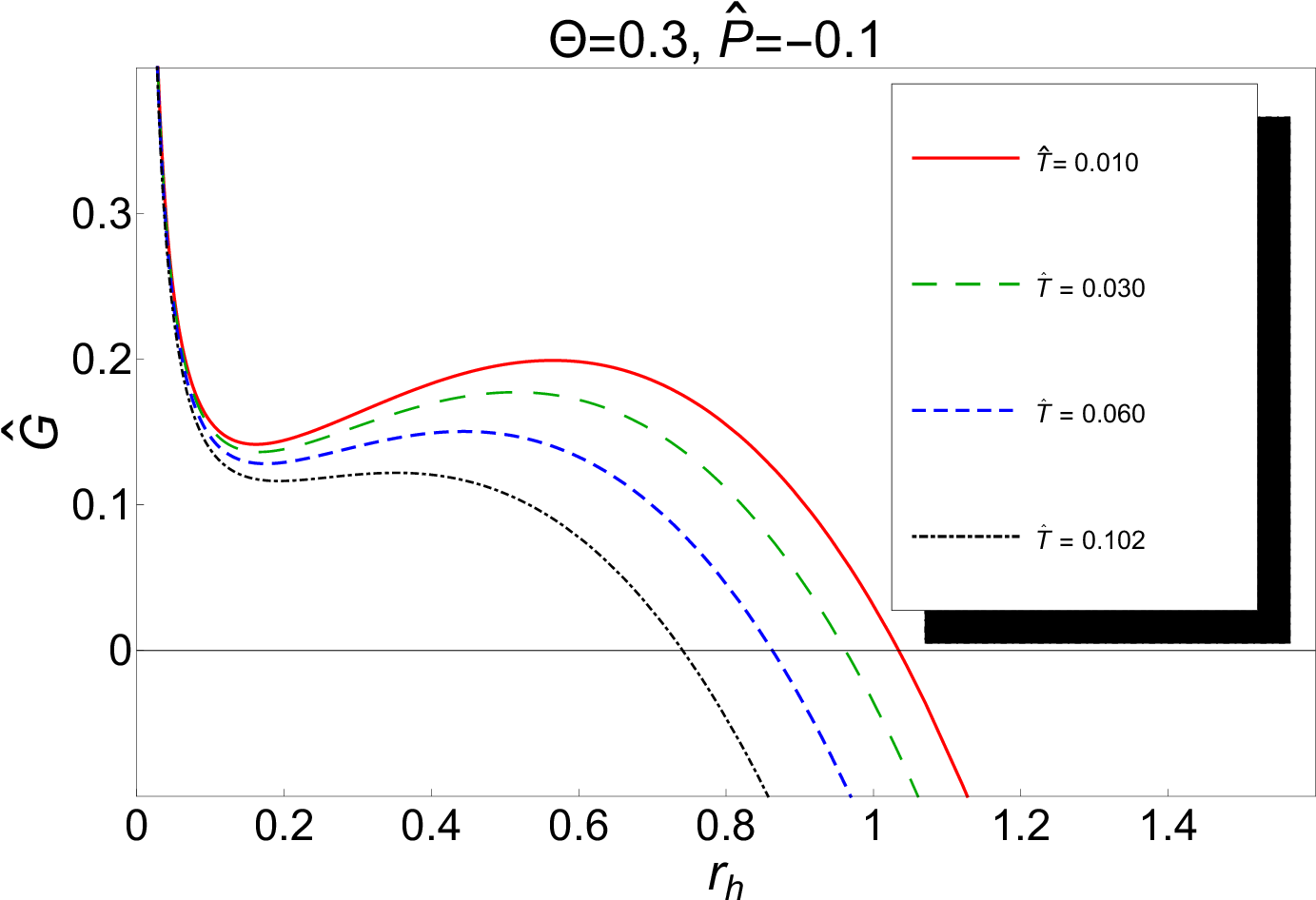}
\caption{Behavior of Gibbs free energy $\hat{G}$ as a function of $r_h$ in NC space-time, for different values of pressure $\hat{P}$ (left panel) and different Hawking temperature $\hat{T}$ (right panel).} \label{fig12}
\end{figure}

We can now explicitly use NC Helmholtz free energy $\hat{F}$ to calculate NC Gibbs free energy $\hat{G}=\hat{F}+\hat{P}\hat{V}$. So that free energy can be obtained with NC internal energy (NC mass of black hole $\hat{m}$) and NC entropy $\hat{F}=\hat{m}-\hat{T}\hat{S}$ (see Eq. \eqref{eqt3.14}). Thus, we can write the NC Gibbs free energy as
\begin{equation}\label{eqt3.21}
\hat{G}=\hat{m}-\hat{T}\hat{S}+\hat{V}\hat{P}\,.
\end{equation}
The influence of pressure on Gibbs free energy is shown in Fig. \ref{fig12}, where we plot the following expression for different $\hat{P}$ and $\hat{T}$
\begin{equation}\label{eqt3.21'}
\hat{G}=\frac{r_h}{2} + \frac{4\pi}{3} r_h^3 \hat{P}  - \pi\, r_h^2  \hat{T} + \Theta^2\left(\frac{1}{8r_h}+ \pi \,r_h \hat{P} -\frac{5\pi}{8}\hat{T}\right).
\end{equation}

In this study we are only interested in the SBH, which has negative pressure for larger ones (see Fig. \ref{fig10}). Hence, in this case we only take the negative pressure. Then the Gibbs free energy exhibits the same behavior as shown in Fig. \ref{fig8}, i.e., there are only two extrema for $\hat{G}$ (see left panel in Fig. \ref{fig12}) for $\hat{T}<\hat{T}^{\max}$ and $-0.10 \leqslant \hat{P}\leqslant 0$ in NC space-time, where the minimum corresponds the stable black hole (smaller SBH) and the maximum corresponds to the unstable one (larger SBH), and these extremum can be determined by solving Eq. \eqref{eqt3.15}. It is worth noting that when the negative pressure gets a value less than ($\hat{P}\ll-0.10$) the extremum of Gibbs free energy disappears. Furthermore, when we change the temperature in the range $0.010 < \hat{T} \leqslant 0.102<\hat{T}^{\max}$ with a fixed pressure ($\hat{P}=-0.10$), then the same Gibbs free energy behavior is observed (see right panel in Fig. \ref{fig12}), i.e., two extrema. Here the depth of the minimum and the maximum of the Gibbs free energy decreases with temperature. As a consequence, the unstable black hole (the larger one) evaporates to the stable one (the small BH), and this agrees with the analysis of the heat capacity, see Fig. \ref{fig7}, and that indicate a phase transition of SBH in the NC spacetime.
\begin{figure}[ht]
\centering
\includegraphics[width=0.485\textwidth]{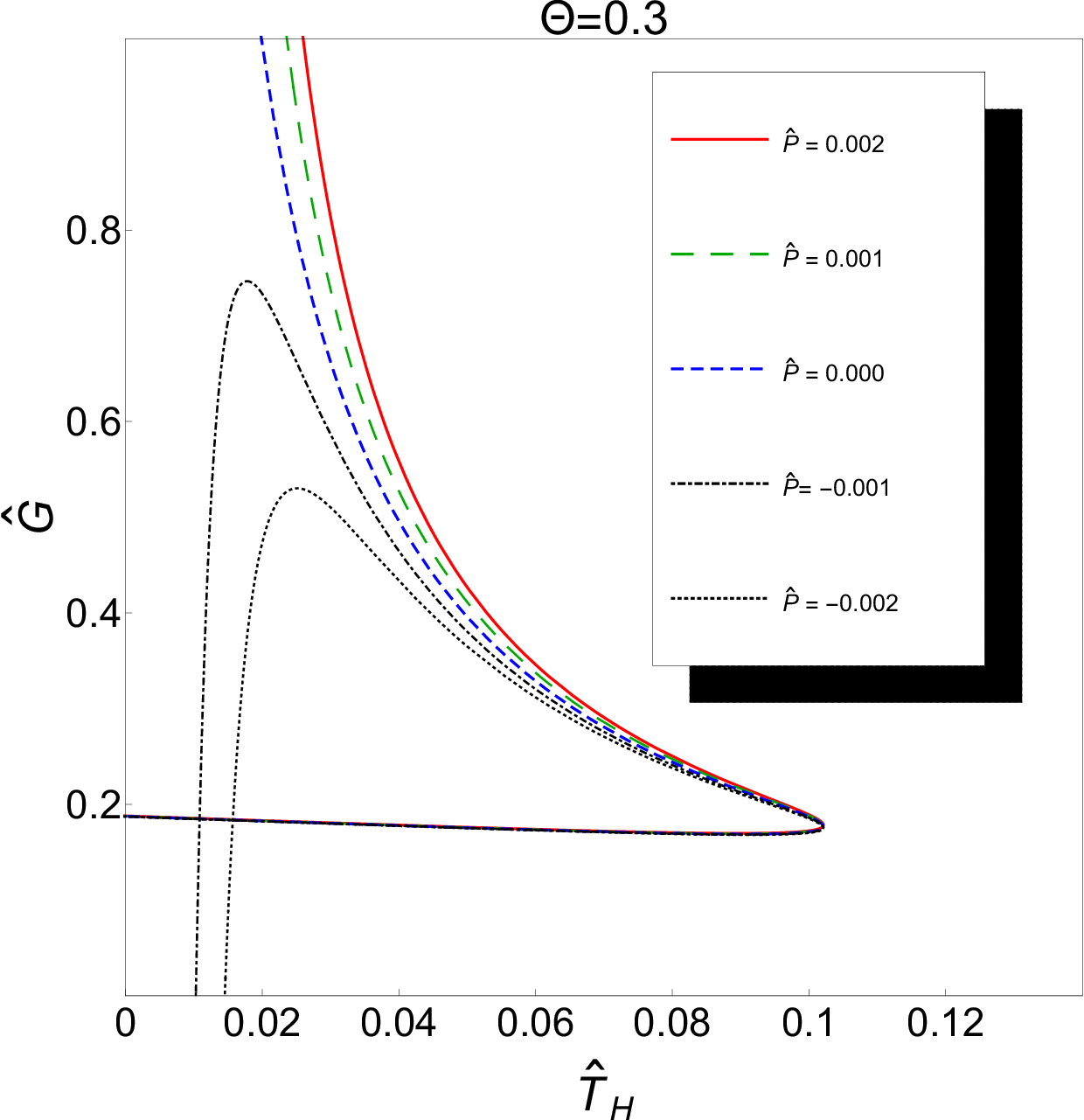}\hfill
\includegraphics[width=0.5\textwidth]{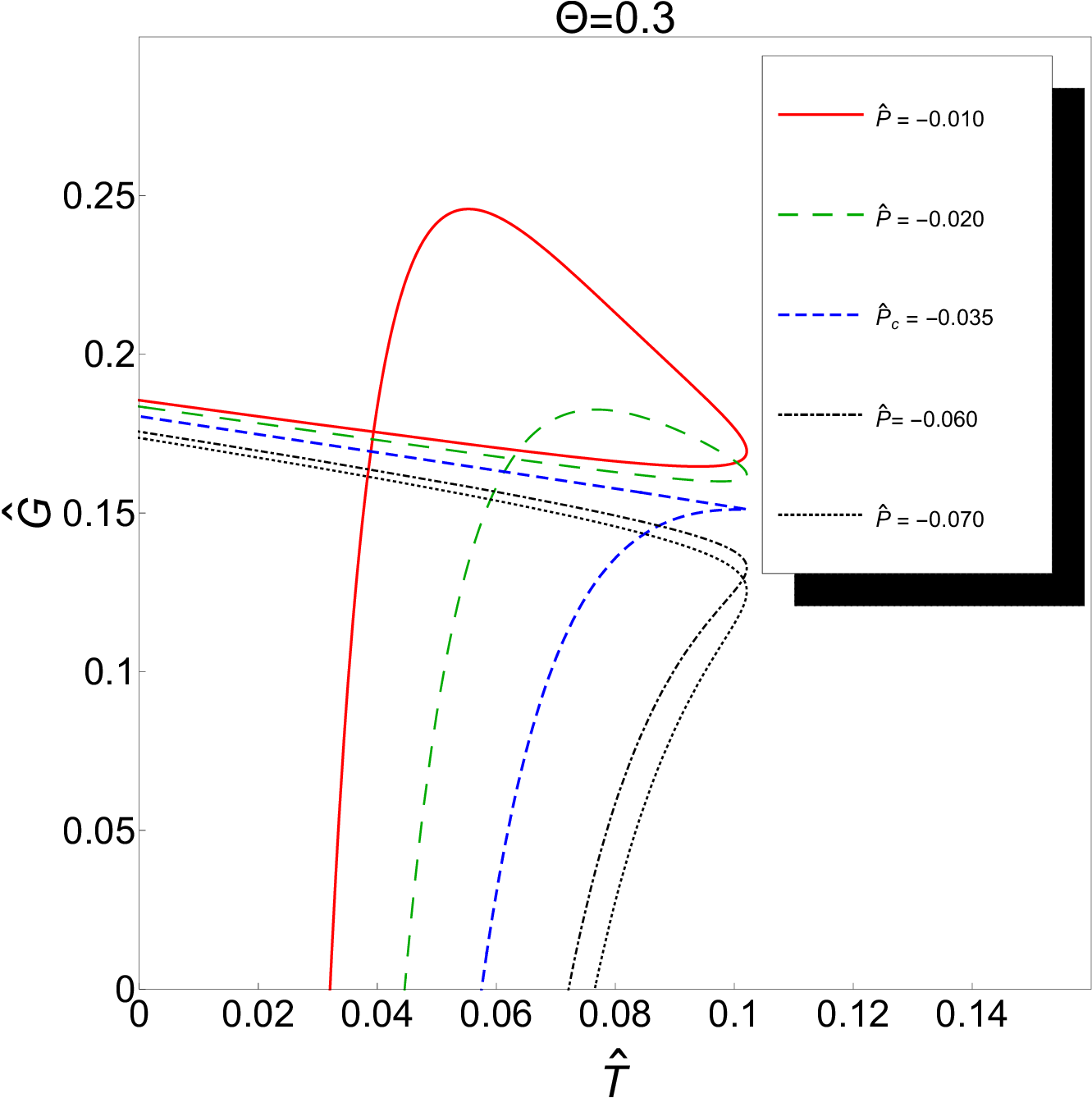}
\caption{Behavior of Gibbs free energy $\hat{G}$ as a function of NC Hawking temperature $\hat{T}$, for different values of NC pressure $\hat{P}$.} \label{fig13}
\end{figure}

The behavior of Gibbs free energy modified with the presence of pressure is shown in Fig. \ref{fig13}, where we plot $\hat{G}$ varying along with NC Hawking temperature for various pressure values. As shown in the left panel, for $ \hat{P} \geqslant 0$, the Gibbs free energy decreases rapidly (from infinity at $\hat{T}\approx0$) as the temperature maximizes, and when it bounces back the Gibbs free energy increases at a slow rate, and that is consistent with the profile "$\hat{F}-\hat{T}$" in Fig. \ref{fig10}. For negative pressure, $\hat{P} \leqslant 0$, the Gibbs free energy increases rapidly as the temperature increases, and when it reaches its maximum, it starts to decrease until the temperature rises to its maximum. Then when the temperature bounces back, the Gibbs free energy increases at a slow rate to coincide with the previous case for the positive pressure. It is worth to note that when we increase the opposite pressure, i.e. $\hat{P}_c < \hat{P} < 0$, we observe a similar behavior to the swallowtail structure as in the literature \cite{hansen,xu,kubiz}, with smoother curves do to the presence of non-commutativity and that does not allow us to interpreted as tow-phases coexistence, where that indicate the absence of the intermediate black hole during his evolution (see Fig. \ref{fig12}). However, at the critical point $\hat{P}_c=-0.035$, where the quasi-swallowtail structure disappears and an inflection point occurs, which is a second-order phase transition, at this point the Hawking-Page like phase transition occurs during the evolution of the NC SBH, where that indicate the unstable black hole (larger one) transit to stable one (small NC SBH), and that consistent with the profile "$\hat{G}-r_h$" in Fig. \ref{fig12} and "$\hat{C}-r_h$" in Fig. \ref{fig7}. The inflection point vanishes when $\hat{P}<\hat{P}_c$, and that means there is one phase transition occurs during the evaporation of the NC SBH at the critical point $\hat{P}=\hat{P}_c$. It is important to mention that, there are some paper that study the effect of quantum gravity on the phase transition of black holes inside a spherical cavity with constant radius $R$ as in Refs. \cite{chen,feng1,md1,feng2,feng3}, and these results show the exact swallowtail structure. However, in our case we show a quasi-swallowtail behavior with smoother curves ($\hat{P}_c < \hat{P} < 0$) without using the spherical cavity and like Hawking-Page phase transition ($\hat{P} =\hat{P}_c$), this behaviors emerges when we study the presence of the black hole pressure in the NC spacetime. 
\begin{figure}[ht]
\centering
\includegraphics[width=0.5\textwidth]{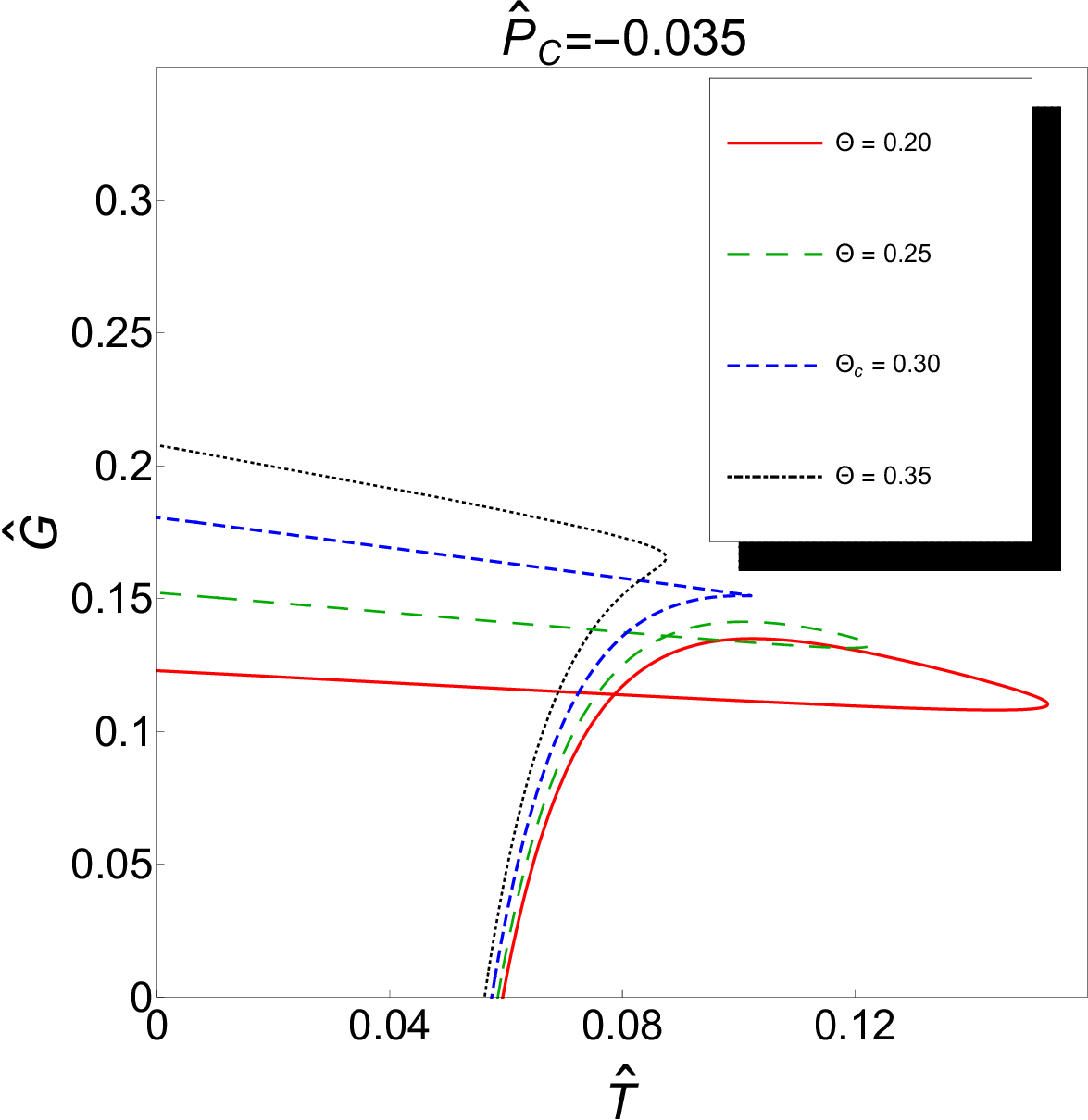}
\caption{Behavior of Gibbs free energy $\hat{G}$ as function of NC Hawking temperature $\hat{T}$, for different values of $\Theta$.} \label{fig14}
\end{figure}

The behavior of Gibbs free energy as a function of NC Hawking temperature $\hat{T}$, for constant pressure $\hat{P}$ and different values of $\Theta$, is shown in Fig. \ref{fig14}. The quasi-swallowtail structure appears with smoother curves for small values of $\Theta$. At a critical point, $\Theta_c=0.30$ the above behavior disappears and an inflection point occurs, a behavior similar to the Hawking-Page phase transition appears, which indicates a phase transition of larger black hole to a smaller one. When $\Theta>\Theta_c$ this inflection point disappears. This behavior is similar to that one obtained in Fig. \ref{fig13} (right panel), but with a difference in the order of the curves. We observe that the NC parameter $\Theta$ plays a role similar to a thermodynamic variable of a black hole.

\section{Conclusions}

In this work, we investigated in detail the thermodynamic proprieties of the SBH in NC gauge theory of gravity. Our results show that the NC corrections affect the event horizon of this black hole $r^{\mathrm{NC}}_h$, and thus, the thermodynamic proprieties of the SBH are also affected. Additionally, non-commutativity shifts the singularity at the origin to a finite radius $r=2m$ \cite{abdellah1}.

As a first step, we corrected the ADM mass which exhibits a new behavior where the SBH has a minimal mass $\hat{m}_0$ in this geometry, and there is no black hole for $\hat{m}<\hat{m}_0$. We also obtained the NC corrections to the Hawking temperature, and found that these corrections remove the divergent behavior appearing in the commutative case and lead to a new maximum temperature that can be reached by a black hole during its evaporation, before quickly falling to zero at a new minimum size $r_h^{\min}$ with a minimum mass $\hat{m}_0$. At this point, the black hole cannot radiate anymore, and there are no more processes of particle-antiparticle creation near the event horizon. The quantum effect of non-commutativity is to cool down the black hole radiation in its final stage (as was shown in Fig. \ref{fig3''}), similarly to the electric charge $Q$ of a Reissner-Nordstr\"{o}m black hole. Then we observed a difference in temperature between the poles and the equator of this black hole in the presence of the observation angle $\theta$, where we noticed that the radiation increases as we move closer to the black hole poles. This can be seen as a jet of radiation from a non-rotating black hole due to the quantum structure of space-time.

There is also a new result that emerges from the NC property of space-time, where the corrections predict a new fundamental length at the Planck scale $\Theta\approx 1.523\times10^{-35}\,\mathrm{m}$. Then we showed that the NC corrections to the black hole entropy are only significant for smaller ones (as we showed in Fig. \ref{fig5}). At the end of the black hole evaporation $r_h \rightarrow 0$, we can still observe a remnant entropy $S^{\Theta}_0$ (see relation \eqref{eq:4.8}). This remnant entropy indicates the presence of a microscopic remnant black hole with a minimal mass $\hat{m}_0$.

We also calculated the correction to the heat capacity. Our results indicate a phase transition of NC SBH, where a larger (unstable) black hole evaporates and transits to a smaller (stable) one. It is worth noting that as the observational angle $\theta$ changes, the critical point of phase transition changes accordingly (the point where $\hat{C}\rightarrow\infty$), so it is not the same in all directions.

To conclude, the non-commutativity gives us a new scenario of black hole evaporation as we summarize below
\begin{itemize}
\item The evaporation of a black hole starts at the equator and goes up to the poles.
\item At the end of the black hole evaporation, we still have a quantum object with a remnant entropy $S^{\Theta}_0$ with minimum mass $\hat{m}_0=0.5\,l_p$ and event horizon $r_h^{min}=l_p$, interpreted as a microscopic remnant black hole and is thermodynamically stable.
\end{itemize}

Finally, a detailed analysis of Helmholtz free energy for NC SBH shows two extremals: one stable (minimum) corresponding to a smaller black hole and another unstable (maximum) for larger ones. This confirms the previous result obtained by the heat capacity analysis. However, for very small horizon radius ($r_h<\Theta$) we found that the Hawking temperature is negative and therefore not physical for global stability. The inclusion of pressure to the NC SBH indicates the same behavior as above, i.e., one minimum (stable) and one maximum (unstable), and these two extremum disappear when $\hat{P}\ll-0.1$. The variation of pressure shows a quasi-swallowtail structure, but the NC geometry makes the curves smoother and shows an inflection point at the critical value $\hat{P}_c$ and a similar behavior to the Hawking-Page phase transition is appears, which means a second-order phase transition. This is consistent with the standard theory of phase transitions. The NC parameter $\Theta$ plays a similar role to a thermodynamical variable. We conclude the existence of a phase transition of NC SBH and predict a new scenario of black hole evaporation and the emergence of a microscopic remnant black hole with a minimal mass $\hat{m}_0$.

\appendix

\acknowledgments

This work is supported by PRFU Research Project B00L02UN050120230003, University of Batna 1, Algeria.

\bibliographystyle{ieeetr}

\bibliography{biblio1}

\end{document}